\documentstyle[12pt]{article}
\title{Falicov-Kimball model and the problem of electronic ferroelectricity}
\author{Pavol Farka\v sovsk\'y\\
Institute  of  Experimental  Physics,  Slovak   Academy   of
Sciences\\
Watsonova 47, 043 53 Ko\v {s}ice, Slovakia}
\date{}
\begin{document}
\baselineskip=24pt
\maketitle

\begin{abstract}

The density matrix renormalization group method is used to examine 
possibilities of electronic ferroelectricity in the spinless Falicov-Kimball
model. The model is studied for a wide range of parameters including 
weak and strong interactions as well as the symmetric and unsymmetric 
case. In all examined cases the $<d^{+}f>$-expectation value vanishes 
for vanishing hybridization $V$, indicating that the spinless 
Falicov-Kimball model does not allow for a ferroelectric ground state 
with a spontaneous polarization. 

\end{abstract}

\newpage

The Falicov-Kimball model~(FKM) has become, since its
introduction~\cite{Falicov} in 1969, one of the most popular examples
of a system of interacting electrons with short-range interactions.
It has been used in the literature to study a great variety of many-body
effects in metals, of which metal-insulator transitions, mixed-valence
phenomena, and charge-density waves are the most common
examples~\cite{Chomski}. In the last few years the FKM is extensively 
studied in connection with the exciting idea of electronic 
ferroelectricity~\cite{P1,P2,Cz,F1,Zl}. It is generally supposed that 
the ferroelectricity in mixed-valent compounds is of purely electronic 
origin, i.e., it results from an electronic phase transition, in contrast 
to the conventional displacive ferroelectricity due to a lattice distortion. 
Since the FKM is probably the simplest model of electronic phase transitions 
in rare-earth  and transition-metal compounds it was natural to start 
to examine possibilities of electronic ferroelectricity just on this model.
An important milestone on this way was the recent work by 
Portengen at al.~\cite{P1,P2}. They studied the FKM with a k-dependent 
hybridization in Hartree-Fock approximation and found, in particular, 
that the Coulomb interaction $U$ between the itinerant $d$-electrons 
and the localized $f$-electrons gives rise a non-vanishing excitonic
$<d^{+}f>$-expectation value even in the limit of vanishing
hybridization $V \rightarrow 0$. As an applied (optical) electrical field
provides for excitations between d- and f-states and thus for a polarization
expectation value $P_{df}=<d_i^{+}f_i>$, the finding of a spontaneous
$P_{df}$ (without hybridization or electric field) has been interpreted as
evidence for electronic ferroelectricity. This result motivated further
theoretical studies of the model. 
Analytical calculations within well controlled approximation (for
$U$ small) performed by Czycholl~\cite{Cz} in infinite dimensions do
not confirm the existence of electronic ferroelectricity. In contrast to 
results obtained by Portengen et al.~\cite{P1,P2} he found that the FKM
in the symmetric case ($E_f=0, n_f=n_d=0.5$) does not allow for a 
ferroelectric ground state with a spontaneous polarization,
i.e., there is no nonvanishing $<d^{+}f>$-expectation value in the
limit of vanishing hybridization.  The same conclusion has been also 
obtained independently by extrapolation of small-cluster 
exact-diagonalization calculations in the one dimension for both intermediate 
and strong interactions~\cite{F1}. In these regions the finite-size
effects are negligible and thus the results can be satisfactory 
extrapolated to the thermodynamic limit. Very recently 
Zlati\'c et. al.~\cite{Zl} analysed the static susceptibility 
for spontaneous polarization in the ordinary FKM ($V=0$) using the exact
solution of the model in infinite dimensions. They found, in particular, 
that the spontaneous hybridization susceptibility diverges at 
$T\to 0$ for $U=0$ and probably for nonzero $U$ too, indicating 
a possibility of a nonzero $<d^+f>$-expectation value in the ground
state.   

In order to shed some light on these controversial results  we performe  
an exhaustive study of the model for a wide range of model parameters.
While in our previous paper the possibilities for the electronic 
ferroelectricity have been studied only in the symmetric case and 
intermediate and strong interactions~\cite{F1} we extend here  
calculations also to the unsymmetric case and small interactions. 
To suppress the finite-size effects in the weak-coupling limit
one has to consider much larger clusters than can be treat numerically
within the exact-diagonalization calculations ($L\sim 12$). 
For this reason we have decided to use in numerical computations the density 
matrix renormalization group (DMRG) method~\cite{White} that allows to treat 
several times larger clusters and still to keep the high accuracy of 
computations.
 

As mentioned above the FKM is based on the coexistence of two
different types of electronic states in a given material:
localized, highly correlated ionic-like states and extended,
uncorrelated, Bloch-like states. It is accepted that
insulator-metal transitions result from a change in the occupation
numbers of these electronic states, which remain themselves basically
unchanged in their character. Taking into account only the
intra-atomic Coulomb interaction between the two types of states,
the Hamiltonian of the spinless FKM with hybridization
can be written as the sum of four terms:

\begin{equation}
H=\sum_{ij}t_{ij}d^+_id_j+U\sum_if^+_if_id^+_id_i+E_f\sum_if^+_if_i+
V\sum_id^+_if_i+h.c.,
\end{equation}
where $f^+_i$, $f_i$ are the creation and annihilation
operators  for an electron in  the localized state at
lattice site $i$ with binding energy $E_f$ and $d^+_i$,
$d_i$ are the creation and annihilation operators
of the itinerant spinless electrons in the $d$-band
Wannier state at site $i$.

The first term of (1) is the kinetic energy corresponding to
quantum-mechanical hopping of the itinerant $d$ electrons
between sites $i$ and $j$. These intersite hopping
transitions are described by the matrix  elements $t_{ij}$,
which are $-t$ if $i$ and $j$ are the nearest neighbors and
zero otherwise (in the following all parameters are measured
in units of $t$). The second term represents the on-site
Coulomb interaction between the $d$-band electrons with density
$n_d=\frac{1}{L}\sum_id^+_id_i$ and the localized
$f$ electrons with density $n_f=\frac{1}{L}\sum_if^+_if_i$,
where $L$ is the number of lattice sites. The third  term stands
for the localized $f$ electrons whose sharp energy level is $E_f$.
The last term represents the hybridization between the itinerant
and localized states.

In order to test the convergence of the DMRG method for the FKM
with hybridization let us first consider the case $U=0$ ($E_f=0$). 
In this case the Hamiltonian (1) can be diagonalized exactly yielding 
the following expressions for the ground-state energy and 
the $P_{df}$ expectation value 

\begin{equation}
E_g=\frac{1}{2L}\sum_{k}(\epsilon_k-(\epsilon_k^2+4V^2)^{1/2}),
\end{equation}

\begin{equation}
P_{df}=-\frac{V}{L}\sum_{k} \frac{1}{(\epsilon_k^2+4V^2)^{1/2}},
\end{equation}
where $\epsilon_k=\frac{1}{L}\sum_{ij}t_{ij}e^{ik(R_i-R_j)}$.
These expressions are valid for arbitrary $L$ and thus they can 
be used directly for comparison with DMRG results obtained on finite 
clusters. Such a comparison shows that the DMRG results 
reproduce the exact results with high accuracy $(\sim 10^{-5})$
even if a relatively small number of states is kept ($m=128$), thereby
the clusters up to 100 sites become accessible for numerical studies.
This is illustrated in Fig.~1a where the difference 
$\Delta E=E_g^{exact}-E_g^{DMRG}$ is plotted as a function of $L$
for $m=128$ and two values of $V$. It is seen that the difference 
$\Delta E$ is smaller than $10^{-5}$ for all clusters of size $L \le 100$. 
One can expect that the finite-size effects will be considerably 
reduced for such cluster sizes and thus the DMRG results
could be satisfactory extrapolated to the thermodynamic limit also for 
weak interactions. Numerical results obtained for the ground-state 
energy $E_g$ (Fig.~1b) and the $P_{df}$ expectation value~(Fig.~2a)
fully confirm this conjecture. It is seen that the finite-size
effects are small for both $E_g$ and $P_{df}$ and so the most important 
feature, and namely the behavior of $P_{df}$ for $V\to 0$ can be
reliably deduced from these results. In all cases the $<d^+f>$-expectation 
value vanishes in the limit $V\to 0$, so there is no spontaneous polarization 
in the one-dimensional spinless FKM with nearest-neighbor hopping~\cite{note} 
for $U=0$. Using the expression $(3)$ one can show immediately 
that the same behavior holds in any dimension.
In Fig.~2b we have displayed numerical results for $P_{df}$ obtained in
the thermodynamic limit for $d=1,2$ and $d=\infty$ (Bethe lattice).
In all cases $P_{df}$ vanishes continuously for $V\to 0$, which
seems to contradict recent results obtained by 
Zlati\'c et. al.~\cite{Zl} for the ordinary FKM ($V=0$) in infinite dimensions.
Indeed, a comparison of our $d=\infty$ results for the $<d^+f>$-expectation
value with $d=\infty$ results obtained by Zlati\'c et. al.~\cite{Zl} for the 
spontaneous hybridization susceptibility $\chi_{hyb}$ leads to very interesting
observation. While the hybridization susceptibility $\chi_{hyb}$ diverges
for $V=0$, the $P_{df}$ expectation value goes to zero in the limit
of vanishing $V$. This can be shown also analytically using the explicit 
expression (Eq.~3) for the $P_{df}$ expectation value.  Replacing 
the summation in Eq.~3 by integration over the energy and taking 
the constant density of states, for simplicity, one can easily show 
that the $P_{df}$ expectation value (in the limit of vanishing $V$)         
behaves like $Vln V$. Thus $P_{df}$ vanishes at $V=0$, although 
$\chi_{hyb}$ diverges at this point. 
This shows that one has to be very careful in doing conclusions about 
the $P_{df}$ behavior from the spontaneous hybridization susceptibility.

The same behavior of the model we expect also for nonzero $U$.
To verify this we perform an exhaustive study of the model for 
a wide range of the model parameters. In addition to the half-filled
band case $n=1$ we examine here in detail also the case $n=0.5, E_f<0$,
where Zlati\'c et. al.~\cite{Zl} found a numerical evidence for a spontaneous
polarization (the divergence of the spontaneous hybridization susceptibility). 

Let us first discuss the DMRG results obtained for $P_{df}$ in the 
half-filled band case and small but finite $U$. This case was not discussed
in our previous paper~\cite{F1} since small-cluster exact-diagonalization 
calculations were not able to give reliable answers due to large 
finite-size effects in this region. As mentioned above within the DMRG
method much large clusters can be treated numerically, thereby the 
finite-size effects are considerably reduced. 
The weak-coupling numerical results for $P_{df}$ are displayed in Fig.~3
for $U=0.5$ and several values of $E_f$. Obviously the one-dimensional 
FKM does not exhibit a ferroelectric ground state with a spontaneous 
polarization in the weak-coupling limit. For all examined values of $E_f$  
the $<d^+f>$-expectation value vanishes for $V\to 0$, and it is 
demonstrated that this result is independent of~$L$. 

The same calculations are performed also for $n=0.5$ and $E_f<0$.
Since the numerical results obtained by Zlati\'c et. al.~\cite{Zl}
show on the possibility of a spontaneous polarization in this
region a very detailed analysis of the model is done in this parameter 
space for both weak and strong interactions.  The typical behaviors 
obtained by numerical simulations are presented in Fig.~4. 
Again there is no spontaneous polarization, neither for weak nor for strong 
interactions. In all cases $P_{df}\to 0$ for $V\to 0$. The finite-size
effects are small for both weak and strong interactions and thus 
these results can be satisfactory extrapolated to the thermodynamic 
limit. 

Thus we can summarize our results. The numerical studies of the spinless FKM 
by density matrix renormalization group method did not confirm a possibility 
of a ferroelectric ground state with a spontaneous polarization in this 
model. In all examined cases (weak and strong interactions, the symmetric
and unsymmetric case) the $P_{df}$ expectation value vanishes for $V\to 0$,
in contrast to what is expected from the behavior of the spontaneous 
hybridization susceptibility~\cite{Zl}. This fact considerably reduces 
(although does not exclude definitively) the possibility for the existence of 
a ferroelectric ground state with a spontaneous polarization in the 
spinless FKM.

\vspace{0.5cm}
This work was supported by the Slovak Grant Agency for Science
under grant No. 2/7021/20.

\newpage

\centerline{\bf Figure Caption}

\vspace{0.5cm}
Fig.~1. (a) The difference $\Delta E=E_g^{exact}-E_g^{DMRG}$ 
as a function of $L$ calculated for $U=0,m=128$ and two different
values of $V$. (b) The ground-state energy as a function of 
hybridization for $U=0$ and $m=128$. Numerical curves corresponding 
to $L=12,24$ and 48 are calculated by DMRG method, while the exact
analytical solution in the thermodynamic limit $L=\infty$ is obtained
from Eq.~2.

\vspace{0.5cm}
Fig.~2. (a) Hybridization dependence of the $<d^+f>$-expectation value
calculated for finite clusters (DMRG results) and in the thermodynamic 
limit (from Eq.~3) for $U=0$. (b) Hybridization dependence of 
the $<d^+f>$-expectation value calculated in the thermodynamic 
limit (from Eq.~3) for $d=1,2$ and $d=\infty$.

\vspace{0.5cm}
Fig.~3. Hybridization dependence of the $<d^+f>$-expectation value
calculated for $U=0.5$ and several values of $E_f$ (DMRG results)
at half-filling $(n=1)$. 
To reveal the finite-size effects the calculations have been performed 
for several finite clusters. 

\vspace{0.5cm}
Fig.~4. Hybridization dependence of the $<d^+f>$-expectation value
calculated for $U=0.5$ (a) and $U=5$ (b) and  several values of 
$E_f$ (DMRG results) at quarter-filling $(n=1/2)$.


\begin{thebibliography}{99}
\bibitem{note} Note that $P_{df}$ can be finite for some special 
(unfortunately, physically not very interesting) types of hopping.
For example,  $P_{df}(V=0)$ is finite in the atomic limit 
$t_{ij}=0$  $(\epsilon_k=0)$ as well as in the case of unconstrained
hopping  $t_{ij}=-t$, for all i,j ($\epsilon_k=-tL\delta_{k,0}$).
\end{thebibliography}
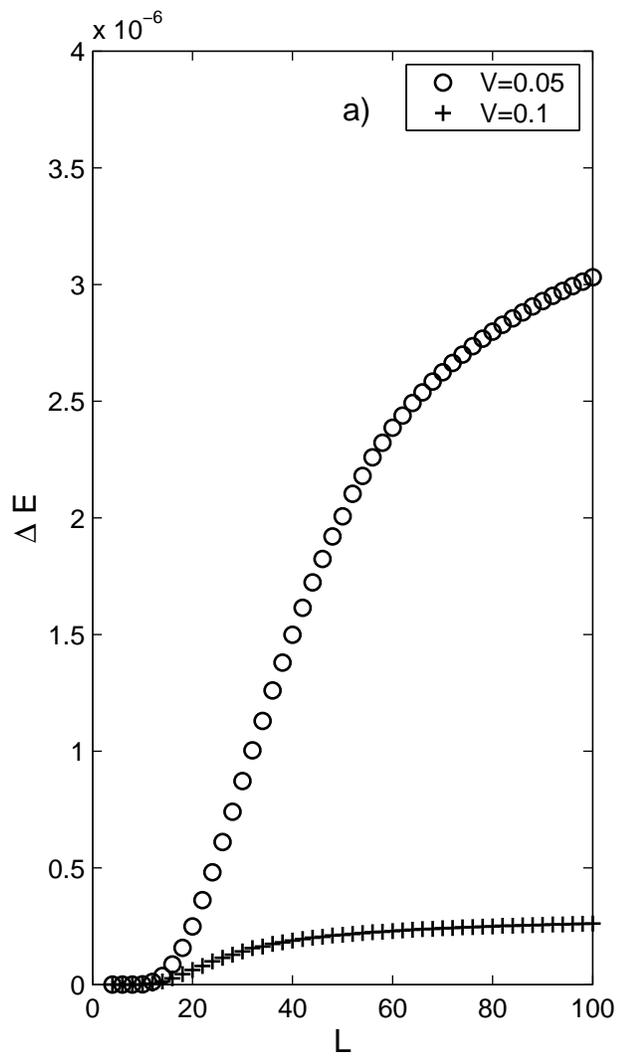
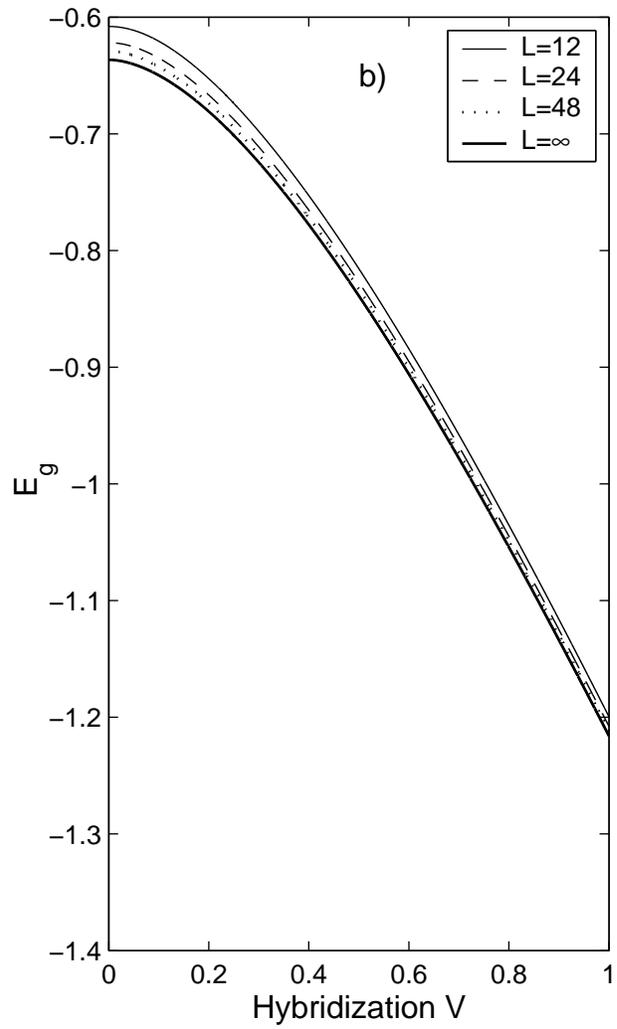
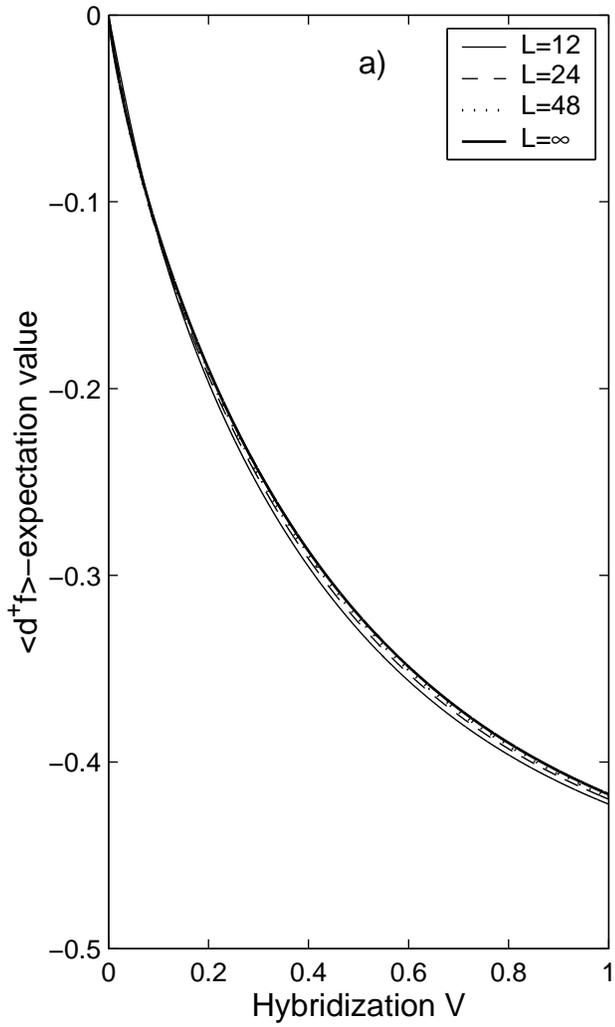
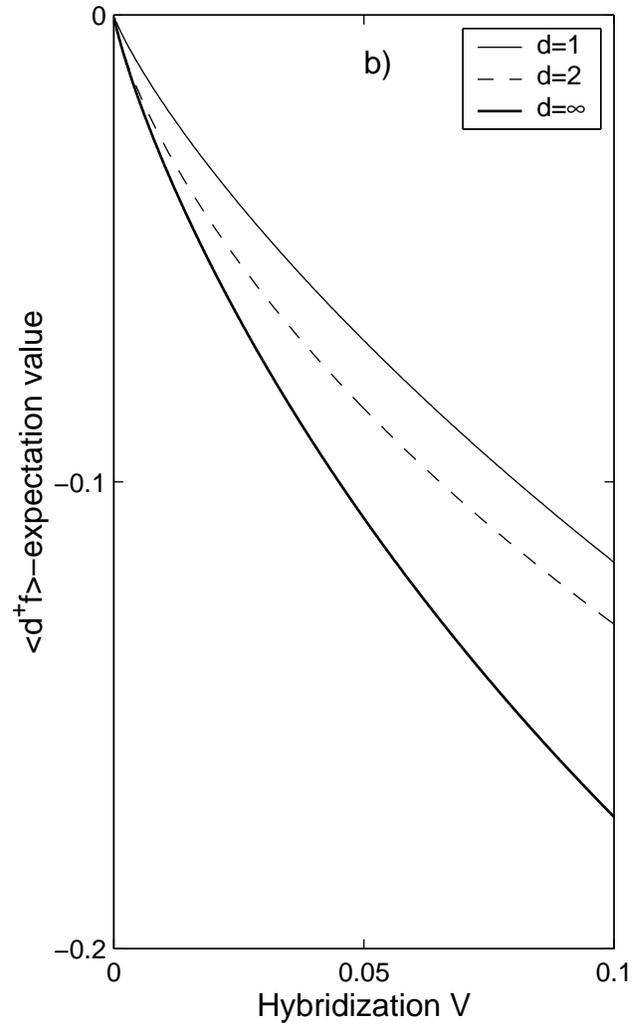
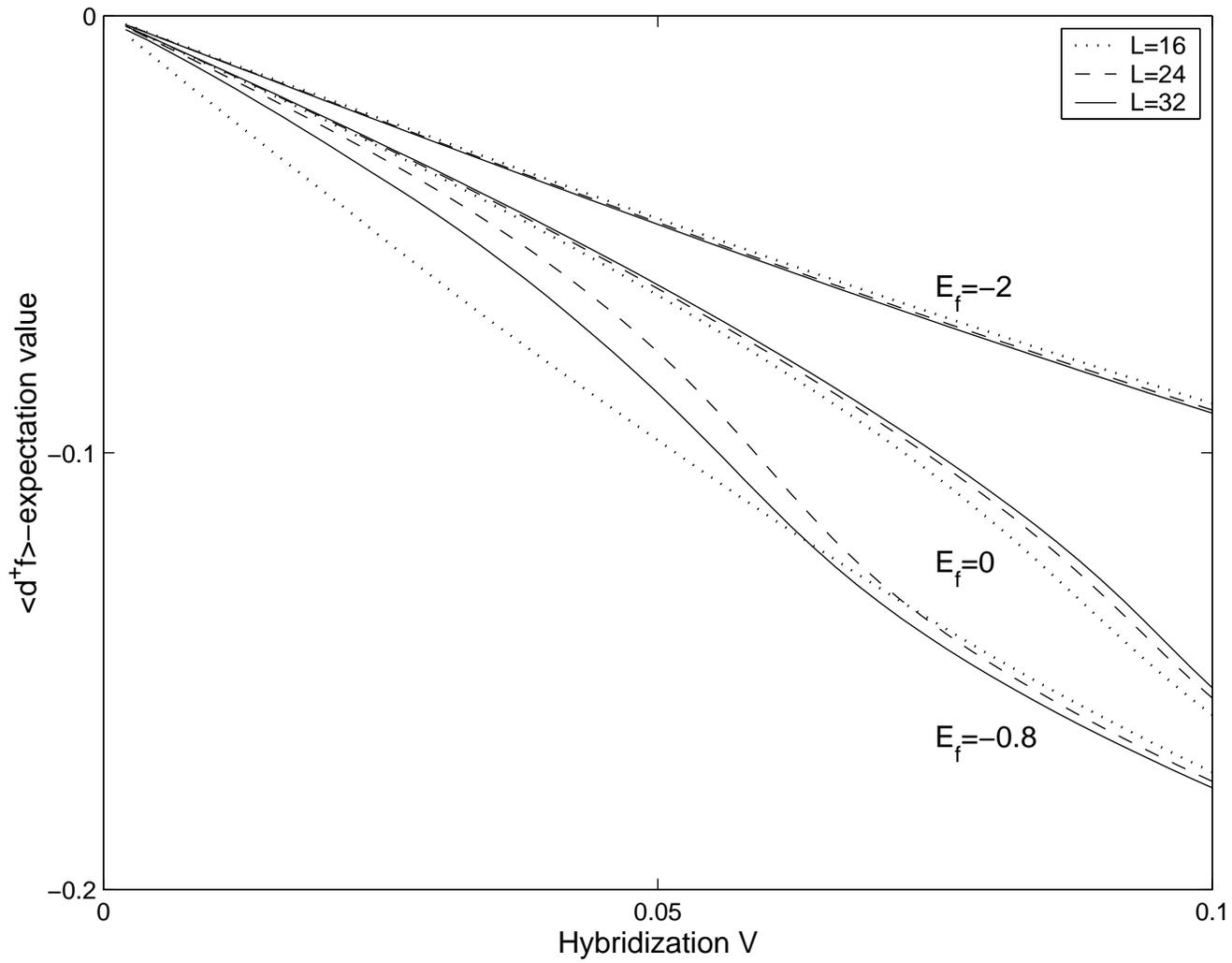
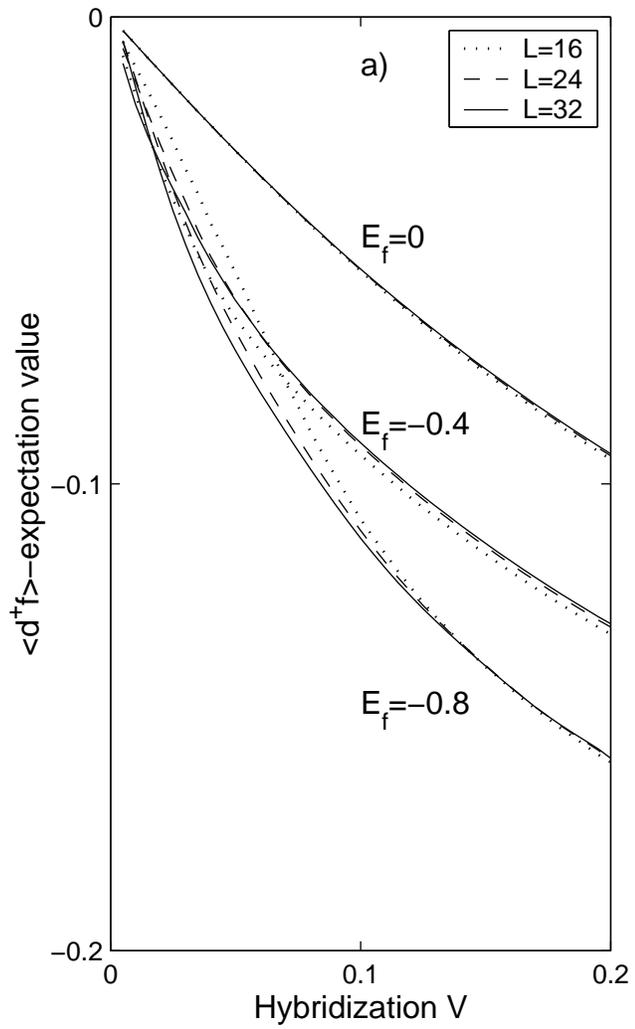
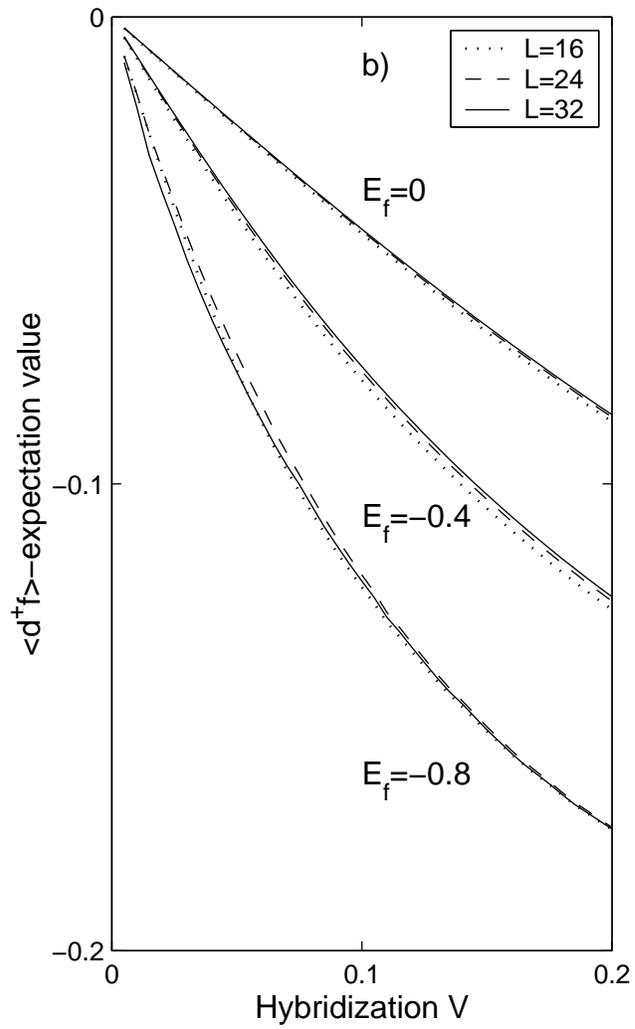
\end{document}